\documentclass[prl,twocolumn,superscriptaddress,showpacs]{revtex4}
\usepackage{graphicx}
\usepackage{amsmath}
\usepackage{amssymb}
\usepackage{dcolumn}
\usepackage{nicefrac}
\newcolumntype{w}[1]{D{.}{.}{#1}}
\newcolumntype{.}{D{x}{}{-1}}

\usepackage{bm}
\usepackage{textcomp}
\usepackage{ulem}
\usepackage{colordvi}
\begin{document}

\title{
Precision Test of Many-Body QED in the Be$^+$ $2p$ Fine Structure Doublet Using Short-Lived Isotopes}

\author{Wilfried N\"ortersh\"auser}
\affiliation{Institut f\"ur Kernphysik, Technische Universit\"at Darmstadt, D-64289 Darmstadt, Germany}

\author{Christopher Geppert}
\affiliation{Institut f\"ur Kernphysik, Technische Universit\"at Darmstadt, D-64289 Darmstadt, Germany}
\affiliation{Institut f\"ur Kernchemie, Johannes Gutenberg-Universit\"at Mainz, D-55128 Mainz, Germany}

\author{Andreas Krieger}
\affiliation{Institut f\"ur Kernphysik, Technische Universit\"at Darmstadt, D-64289 Darmstadt, Germany}
\affiliation{Institut f\"ur Kernchemie, Johannes Gutenberg-Universit\"at Mainz, D-55128 Mainz, Germany}

\author{Krzysztof Pachucki}
\affiliation{Faculty of Physics, University of Warsaw, Pasteura 5, 02-093 Warsaw, Poland}

\author{Mariusz Puchalski}
\affiliation{Faculty of Physics, University of Warsaw, Pasteura 5, 02-093 Warsaw, Poland}
\affiliation{Faculty of Chemistry, Adam Mickiewicz University, Umultowska 89b, 61-614 Pozna{\'n}, Poland}

\author{Klaus Blaum}
\affiliation{Max-Planck-Institut f\"ur Kernphysik, D-69117 Heidelberg, Germany}

\author{Mark L.~Bissell}
\affiliation{Instituut voor Kern- en Stralingsfysica, KU Leuven, B-3001 Leuven, Belgium}

\author{Nadja Fr\"ommgen}
\affiliation{Institut f\"ur Kernchemie, Johannes Gutenberg-Universit\"at Mainz, D-55128 Mainz, Germany}

\author{Michael Hammen}
\affiliation{Institut f\"ur Kernchemie, Johannes Gutenberg-Universit\"at Mainz, D-55128 Mainz, Germany}

\author{Magdalena Kowalska}
\affiliation{CERN, Physics Department, CH-1211 Geneva 23, Switzerland}

\author{J\"org Kr\"amer}
\affiliation{Institut f\"ur Kernphysik, Technische Universit\"at Darmstadt, D-64289 Darmstadt, Germany}

\author{Kim Kreim}
\affiliation{Max-Planck-Institut f\"ur Kernphysik, D-69117 Heidelberg, Germany}

\author{Rainer Neugart}
\affiliation{Institut f\"ur Kernchemie, Johannes Gutenberg-Universit\"at Mainz, D-55128 Mainz, Germany}

\author{Gerda Neyens}
\affiliation{Instituut voor Kern- en Stralingsfysica, KU Leuven, B-3001 Leuven, Belgium}

\author{Rodolfo S\'anchez}
\affiliation{GSI Helmholtzzentrum f\"ur Schwerionenforschung GmbH, D-64291 Darmstadt, Germany}

\author{Deyan T.~Yordanov}
\affiliation{CERN, Physics Department, CH-1211 Geneva 23, Switzerland}

\date{\today}
\pacs{32.10.Fn 31.15.ac 31.15.aj 31.30.jf}

\newcommand{\fm}{\ensuremath{\textrm{fm}}}

\begin{abstract}
Absolute transition frequencies of the $2s\;^2\mathrm{S}_{\nicefrac{1}{2}} \rightarrow 2p\;^2\mathrm{P}_{\nicefrac{1}{2},\, \nicefrac{3}{2}}$ transitions in Be$^+$ were measured for the isotopes $^{7,9-12}$Be. The fine structure splitting of the $2p$ state and its isotope dependence are extracted and compared to results of \textit{ab initio} calculations using explicitly correlated basis functions, including relativistic and quantum electrodynamics effects at the order of $m \alpha^6$ and  $m \alpha^7 \ln \alpha$. Accuracy has been improved in both the theory and experiment by 2 orders of magnitude, and good agreement is observed. This represents one of the most accurate tests of quantum electrodynamics for many-electron systems, being insensitive to nuclear uncertainties. 
\end{abstract}

\maketitle
    
%\section{Introduction}

Fine structure splittings in two-electron atoms have attracted much interest as a test of bound-state QED for a long time. Not only helium \cite{Borbely2009} but also heavier heliumlike systems up to fluorine F$^{7+}$ have been studied by using laser spectroscopy \cite{Din1991,Tho1998,Mye1999}. While the helium fine structure was calculated up to the order of $m\alpha^7$ and currently serves as one of the most precise QED tests in two-electron systems \cite{Pac2010}, the extension of such calculations to three-electron systems proved to be much harder. 
The main reason is the considerably more difficult application of the three-electron computational methods with explicitly correlated functions as compared to the two-electron ones. Nevertheless, it has been recently possible to perform the complete calculation of $m \alpha^6$ and $m \alpha^7 \ln \alpha$ contributions to the lithium fine structure \cite{Puc2014} leading to the most accurate QED test with lithium atoms.

Measurements of the $2p$ fine structure splitting in light three-electron systems are limited in accuracy for isotopes with nonzero nuclear spin due to the unresolved hyperfine structure (hfs) in the $2p_{\nicefrac{3}{2}}$ level. This has been the reason for the fluctuating fine structure splittings in lithium \cite{Nob2006,Noe2011} being reported for a long time and turned out to be caused by quantum interference effects in the observation of the unresolved resonance lines \cite{San2011}. Once this issue had been resolved experimentally, good agreement with \textit{ab initio} calculations was obtained \cite{Bro2013}. 

With increasing $Z$, relativistic and QED contributions grow in size and studying such systems allows for further tests of bound-state QED. However, only for the lightest systems, the nonrelativistic QED (NRQED) perturbative approach can be used. Already in the mid-$Z$ region, e.g.\ Si$^{11+}$, relativistic effects in the electron-nucleus interaction must be accounted for exactly by solving the Dirac equation being correct to all orders in the electron-nucleus interaction $\alpha Z$. In this nonperturbative approach an explicit treatment of electron correlations is not possible anymore. Instead, the interelectron interaction is expanded in a perturbation series of $1/Z$ and $\alpha$ \cite{Volotka2014}. Hence, these tests in light and heavy ion systems probe QED at different values of the field strength and are thus complementary.
%With increasing $Z$, relativistic and QED contributions grow in size and studying such systems allow for further tests of bound-state QED. However, only for the lightest systems, the nonrelativistic QED (NRQED) perturbative approach can be used. Already in the mid-$Z$ region, e.g.\ Si$^{11+}$, relativistic effects in the electron-nucleus interaction must be accounted for exactly by solving the Dirac equation being correct to all orders in the electron-nucleus interaction $\alpha Z$. In this non-perturbative approach an explicit treatment of electron correlations is not possible anymore. Instead, the interelectron interaction is expanded in a perturbation series of $1/Z$ and $\alpha$, respectively\cite{Volotka2014}. 
%Hence, tests of QED and relativistic corrections in light and heavy ion systems are complementary since they probe QED at different values of the field strength.
The most accurate tests of QED in heavier three-electron systems are those of the $2s_{1/2}\rightarrow 2p_{1/2}$ transition energy in lithiumlike uranium \cite{Beiersdorfer2005} and the $g$ factor in Si$^{11+}$ \cite{Wagner2013}. 
%The comparison of the hfs in hydrogen- and lithium-like Bi \cite{Lochmann14,Ullmann15} provides another test, albeit at lower precision yet. In general these heavy ion tests address strong field QED and are complementary to those in light ions, where weak field QED is combined with electron correlations.

For further tests on low-$Z$ ions, the Be$^+$ and B$^{2+}$ ions are suitable candidates, since their transition wavelengths at 313\,nm and 205\,nm, respectively, are still accessible by using cw lasers with second-harmonic or fourth-harmonic generation. The most accurate measurement of the splitting in Be$^+$ ions was performed in a Penning trap with a precision of about 60\,MHz \cite{bollinger1985}. Unfortunately, there is no stable isotope with zero nuclear spin below $^{12}$C forming a three-electron system. However, radioactive ion beam facilities can provide the isotopes $^{10}$Be and $^{12}$Be with lifetimes of $1.6 \times 10^{6}$\,a and 20\,ms, respectively. These have zero nuclear spin and are thus ideal candidates for an accurate determination of the fine structure splitting in a three-electron, $Z=4$ system. Other advantages of the even-even isotopes are the absence of quantum interference effects that lead to problems in the case of lithium isotopes and hyperfine-induced fine structure mixing that can also affect the splitting magnitude. 

In this Letter, we report on experimental and theoretical results on the total transition frequencies and $2p_{\nicefrac{1}{2},\,\nicefrac{3}{2}}$ fine structure splittings in $^{7,9-12}$Be$^+$. The experimental accuracy is improved by 2 orders of magnitude for the stable isotope $^9$Be and the splittings in the radioactive isotopes are reported for the first time. They are all obtained from frequency measurements in the $2s_{\nicefrac{1}{2}}\rightarrow 2p_{\nicefrac{1}{2}}$ ('D1') and the $2s_{\nicefrac{1}{2}}\rightarrow 2p_{\nicefrac{3}{2}}$ ('D2') transitions using a sophisticated variant of (on-line) collinear laser spectroscopy \cite{Noe2009}. Moreover, they yield the variation of the fine structure splitting along the chain of isotopes, the so-called splitting isotope shift (SIS). This differential observable can be extracted with high accuracy from the calculations since the mass-independent relativistic and QED contributions cancel out. The SIS provides also a valuable consistency check of the experimental results \cite{lit_fine_yd}. 
Finally, we take advantage of the fact that the fine structure splitting has been measured on a chain of isotopes, among them two spin-zero isotopes without hfs. The higher accuracy of these measurements, being insensitive to nuclear structure corrections, are transferred to the stable isotope $^9$Be using the calculated SIS. This procedure reduces the splitting uncertainty in $^9$Be by another factor of 4 and represents now 
together with the fine structure splitting in lithium \cite{Bro2013,Puc2014} the highest-precision test of relativistic many-body theory for light many-electron systems.   
    
%\section{Experimental}

Absolute frequency measurements on $^{7,9-12}$Be in a fast ion beam ($\beta=\upsilon/c = 3\times 10^{-3}$) were performed by applying the frequency-comb-based simultaneous collinear-anticollinear spectroscopy technique \cite{Noe2009}. Unlike the standard collinear laser spectroscopy approach, this technique allows one to extract also total transition frequencies with high accuracy. This is based on the simple relation from special relativity for the rest-frame transition frequency %$\nu_0$  
%\begin{equation}
	$\nu^{2}_{0} = \nu_{a} \cdot \nu_{c}$
%	\label{Eq:nu0}
%\end{equation}
that has recently been tested to parts-per-billion accuracy \cite{Bot2014}. The frequencies $\nu_{a}$ and $\nu_{c}$ are the laser frequencies measured in the laboratory system at which resonant excitation of the ion beam is observed with anticollinear and collinear laser beams, respectively. A similar approach was used in the past to determine fine structure splittings in He-like ions of the second row of the periodic table to test QED calculations \cite{Din1991,Tho1998,Mye1999} and to calibrate acceleration voltages in on-line collinear spectroscopy \cite{Din1989}. The availability of frequency combs \cite{Rei1999} facilitates this technique and the measurements performed here are more than an order of magnitude more precise than those reported before on He-like systems. 

In order to extract the fine structure splitting, the optical transition frequencies of the D1 and D2 lines at about 313~nm were determined. The data presented here were collected in two beam times (run I, run II) at the radioactive beam facility ISOLDE/CERN. 
%Details of the experimental setups are presented in \cite{Noe2009,Kri2012} and only a brief description will be given here. 
The different isotopes were produced in fragmentation reactions induced by 1.4-GeV protons impinging on a uranium carbide target, laser ionized and delivered with beam energies of about 50\,keV to the collinear laser spectroscopy setup COLLAPS. 
Here, the ion beam was superimposed with copropagating and counterpropagating laser beams behind a 10$^\circ$ electrostatic deflector with an angular deviation of less than 0.5\,mrad. 
%The following paragraph can be shortened 
%Laser light was produced with two continuous-wave dye lasers operating at the fundamental wavelengths of 624\,nm and 628\,nm for collinear and anticollinear excitation, respectively. After fiber transport to the spectroscopy station, the uv light was produced in two second-harmonic generation cavities and superimposed with the ion beam. Absolute transition frequencies in the laboratory system were determined by stabilizing one of the two lasers to various transition lines in the spectrum of molecular iodine and the other one directly to a frequency comb. The transition frequency of the iodine-stabilized laser was regularly checked with the frequency comb, to become independent from all effects that could lead to a shift of the locking point. No detectable shift could be observed during the whole beamtime of about one week.
%Possible shortened version:
Laser light was produced with two continuous-wave dye lasers operating at the fundamental wavelengths of 624\,nm and 628\,nm for collinear and anticollinear excitation, respectively. They are both referenced to a frequency comb as described in Refs.\ \cite{Noe2009,Kri2012}and the second harmonic was generated after fiber transport in two external cavities at the beam line.  

Resonance fluorescence was observed in the optical detection region, at which voltages up to $\pm 10$\,kV can be applied for Doppler tuning. For $^{12}$Be, an ion-photon coincidence detection was used for background rejection of scattered laser light when detecting the signal with the low production rate of about 8000 ions/s \cite{Kri2012}. 
The hfs in the $2s~^2{\rm{S}}_{\nicefrac{1}{2}} \rightarrow 2p~^2{\rm{P}}_{\nicefrac{1}{2},\,\nicefrac{3}{2}}$ transitions was fitted by using the Casimir formula with the hyperfine coefficients $A$ and $B$. The even isotopes $^{10,12}$Be do not exhibit hfs and have been used to study the experimental lineshape. 

\begin{figure}[tbp]
	\includegraphics[width=\columnwidth]{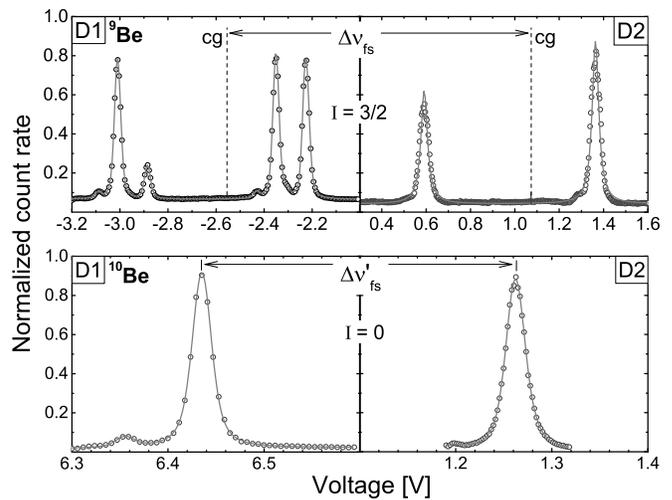}
	\caption{Fluorescence spectra of $^{9}$Be (top row) and $^{10}$Be (bottom row) in the $2s_{\nicefrac{1}{2}}\rightarrow 2p_{\nicefrac{1}{2}}$ (left,\,'D1') and the $2s_{\nicefrac{1}{2}}\rightarrow 2p_{\nicefrac{3}{2}}$ (right,\,'D2') transition as a function of the Doppler-tuning voltage applied to the high-voltage amplifier. A two-component Voigt profile was fitted for each hyperfine component to account for the small satellite peak caused by inelastic collisions. The distance corresponding to the $2p$ fine structure splitting $\Delta \nu_\mathrm{fs}$ is indicated.}
\label{fig:1}
\end{figure}

%\squeezetable
\begin{table*}
\begin{center}
\caption{Absolute transition frequencies $\nu_0$ for the beryllium isotopes under investigation obtained in run I and II. The first uncertainties are statistical and the second ones are the total uncertainty. In the other columns the fine structure splittings $\Delta \nu_{\mathrm{fs}}$, the experimental and theoretical \cite{Puc2009} splitting isotope shifts $\delta \nu_{\mathrm{sis}}$, and the transferred fine structure splittings $\Delta \nu_{\mathrm{fs,^ABe\rightarrow\,^9Be}}$ for $^9$Be based on the measured splittings in the radioactive isotopes according to Eq.\,(\ref{eq:fs-transf}) are listed. The bottom row shows the splitting isotopes shift between the two even isotopes $^{10}$Be and $^{12}$Be. Evaluation of $\delta \nu^{A,9}_{\mathrm{sis}}$ and $\Delta \nu_{\mathrm{fs,^ABe\rightarrow\,^9Be}}$ for $^{7,9,11}$Be after run II required information from run I since D2 lines of these isotopes were not measured in run II. All values are in MHz.}
%Note, that $^{7}$Be was not investigated in the measurement campaign 2010 anymore. Hence, the frequency was overtaken from the 2008 data, but the total uncertainty was re-calculated in respect to the investigations of the photon recoil effect which was overestimated with more than 1~MHz.}
\label{tab:expfrequency}
\vspace{2mm}
\begin{tabular}{c c r@{.}l r@{.}l r@{.}l r@{.}l r@{.}l r@{.}l}
\hline\hline
Isotope & Run & \multicolumn{2}{c}{$\nu_0$} &\multicolumn{2}{c}{$\nu_0$} &\multicolumn{2}{c}{$\Delta \nu_\mathrm{fs}$} & \multicolumn{4}{c}{$\delta \nu^{A,9}_{\mathrm{sis}}$} & \multicolumn{2}{c}{$\Delta \nu_{\mathrm{fs,^ABe\rightarrow\,^9Be}}$} \\
& & \multicolumn{2}{c}{D1}      &\multicolumn{2}{c}{D2}      &\multicolumn{2}{c}{D2--D1}                   & \multicolumn{2}{c}{Exp} & \multicolumn{2}{c}{Theory}\\%& /MHz & /MHz & /MHz & /MHz & /fm$^2$ & /fm \\
\hline
~$^{7}$Be & I  & 957\,150\,316&2 (8) (9)    & 957\,347\,374&5  (9) (11)  & 197\,058&4 (14)   &   5&1 (22)$^a$ \\
~$^{7}$Be & II & \multicolumn{2}{c}{--}      & \multicolumn{2}{c}{--}     & \multicolumn{2}{c}{--} &4&9 (21)   & 6&036(1)    & 197\,064&4 (14) \\
~$^{9}$Be & I  & 957\,199\,552&9 (8) (10)   & 957\,396\,616&6  (14)(15) & 197\,063&7 (20)		\\
~$^{9}$Be & II & 957\,199\,553&40 (12)(52)  & \multicolumn{2}{c}{--}		  & 197\,063&2 (16)   &   0&0 			    & 0&0			 & 197\,063&2 (16)	\\
~$^{9}$Be &$^b$& \multicolumn{2}{l}{957\,199\,652 (120)}     & 	\multicolumn{2}{l}{957\,396\,802 (135)}    & \multicolumn{2}{l}{197\,150 (64)} \\

$^{10}$Be & I	 & 957\,216\,876&9  (14)(15)  & 957\,413\,943&9 (8) (10)  & 197\,067&0 (23) 	& --3&5 (24)$^a$	 \\
$^{10}$Be & II & 957\,216\,876&84 (42)(66)  & 957\,413\,942&17 (10)(55) & 197\,065&3 (9) 	  & --2&1 (18)      & --2&096(1)  & 197\,063&2 (9) \\

$^{11}$Be	& I	 & 957\,231\,118&1  (11)(12)  & 957\,428\,185&2 (14)(16) & 197\,067&1 (23)   & --3&6 (25)$^a$	\\
$^{11}$Be	& II & 957\,231\,118&11 (10)(52)  & \multicolumn{2}{c}{--}		  & 197\,067&1 (17)   & --3&9 (23)      & --3&965(1)  & 197\,063&1 (17) \\

$^{12}$Be & II & 957\,242\,944&86 (33)(61)  & 957\,440\,013&60 (28)(58) & 197\,068&7 (9) 	  & --5&5 (18) 		  & --5&300(1)	 & 197\,063&4 (9) \\
$^{12-10}$Be & II &	\multicolumn{2}{c}{--}  & \multicolumn{2}{c}{--}		& \multicolumn{2}{r}{$\delta \nu^{12,10}_{\mathrm{sis}}=$}	& --3&4 (6)  	 & --3&203 \\
\hline\hline\\
\multicolumn{9}{l}{$^a$ Zakova \textit{et al.} \cite{Zak2010}, $^b$ Bollinger \textit{et al.} \cite{bollinger1985}}\\
\end{tabular}
\end{center}
\end{table*}

The individual lines, shown in Fig.\,\ref{fig:1}, each exhibit a satellite peak occurring at higher beam energy. It is caused by inelastic collisions during the transport through the beam line, which can lead to excitation of the $2p$ state transferring motional energy into the atomic system which subsequently decays to the ground state. Hence, a two-peak structure consisting of a main peak and a satellite shifted by 4\,V, both with Voigt profiles, is fitted for each hyperfine component. The Lorentzian linewidth of the Voigt profile was kept fixed at the natural linewidth of 19.64\,MHz. The Doppler width, the interval factors $A$ and $B$, the ratio of the main to satellite peak intensities, the intensities of the main peaks and the center of gravity (cg) were free parameters for $\chi^2$ minimization. From the fitting a full width at half maximum of about 40\,MHz was obtained, resulting from a residual Doppler width of $\approx 30$\,MHz.

Fitting independently the collinear and the anticollinear spectra, we obtain the cg of the hfs for the collinear scan $\nu_c$ and the anticollinear scan $\nu_a$. These were then used to calculate the absolute rest-frame transition frequency $\nu_0$ %
%%instead of the following paragraph we can shorten this to 
%according to Eq.\,(\ref{Eq:nu0}) 
and adding the recoil-correction term $\delta \nu_{\rm{rec}} = h \nu^2_{\rm{photon}}/Mc^2$, which corresponds to the shift required to ensure energy and momentum conservation during the absorption process of the photon. 
%For shortening remove from here
%However, Eq.\,(\ref{Eq:nu0}) requires that the cg of both spectra appear exactly at the same voltage. Once both resonances were found, the laser frequency for anticollinear excitation could be shifted accordingly due to the direct lock to the frequency comb. Thus, it was possible to keep the deviation between the corresponding centers $\Delta U \lesssim 0.5$\,V. This small remaining shift was accounted for by the correction 
%\begin{equation}
	%\label{eq:diffdopplertransition}
	%\nu_0 = \sqrt{\left(\nu_c - \frac{\partial \nu}{\partial U} \cdot \Delta U \right) \cdot \nu_{a}} - \delta \nu_{\rm{rec}}.
%\end{equation}
%with the differential Doppler shift $\partial \nu / \partial U \approx e \nu /\sqrt{2eU Mc^2}$. The additional term $\delta \nu_{\rm{rec}} = h \nu^2_{\rm{photon}}/Mc^2$ 
%denotes the recoil correction term which must be included due to energy and momentum conservation during the absorption/emission process of photons.
%% to here
It contributes with about 200\,kHz to the absolute transition frequency. 

The transition frequencies of the D1 and D2 lines from both beam times are summarized in Table\,\ref{tab:expfrequency}. Only measurements of the stable isotope $^9$Be were reported previously \cite{bollinger1985} but had two orders of magnitude less accuracy. Our statistical uncertainty 
%obtained for our measurements 
%from the fitting 
%and error propagation through 
%long version
%Eq.\,(\ref{eq:diffdopplertransition})
%%in shortened version
%Eq.\,(\ref{Eq:nu0})
%
is indicated in the first parentheses, while the total uncertainty
%, including a systematic uncertainty of 510\,kHz added in quadrature, 
is listed in the second ones: A systematic uncertainty of 510\,kHz was added in quadrature,
arising from a possible misalignment between ion and laser beams or both laser beams (500\,kHz), uncertainty of the Rb clock frequency used for the frequency comb (40\,kHz), and a small recoil contribution due to multiple scattering of photons (100\,kHz). Uncertainties for the misalignment and the recoil were determined experimentally by measuring the observed shifts with intended misalignment and by studying the power dependence of the resonance position, respectively. It is obvious from Fig.\,\ref{fig:1} that the determination of the fine structure splitting in the even isotopes $^{10,12}$Be is much easier than in the odd isotopes, where especially the cg in the D2 line is less accurate due to the unresolved hfs. Note that each of the two peaks consists actually of three components. Contrarily, in the even isotopes $\Delta \nu_\mathrm{fs}$ is just given by the peak distance. 

Based on the experimental transition frequencies, the fine structure splitting $\Delta \nu_\mathrm{fs}$ and the SIS $\delta \nu^{A,9}_\mathrm{sis}=\Delta \nu_\mathrm{fs} (^{9}\mathrm{Be}) - \Delta \nu_\mathrm{fs} (^{A}\mathrm{Be})$ were calculated and are included in Tab.\,\ref{tab:expfrequency}. 
%For the uncertainty estimation, 
The total uncertainties of the transition frequencies were added in quadrature since the dominant part (beam alignment) is uncorrelated between the two beam times. For measurements that were both taken during one beam time, this might lead to an overestimation of the total uncertainty. The fine structure splitting in $^9$Be can be compared with accurate theoretical calculations briefly presented in the following.

\begin{figure}[tbp]
	\includegraphics[trim=10mm 5mm 30mm 10mm, width=\columnwidth]{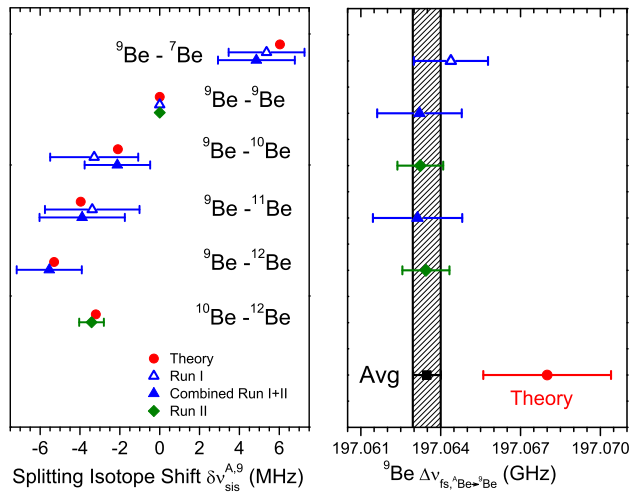}
	\caption{Experimental results and comparison with theory for the splitting isotope shift (left) and for the fine structure splitting of all isotopes transferred to $^9$Be (right). The legend belongs to both graphs. The SIS data were clearly improved in run II compared to the previously reported data \cite{Zak2010}. Combining all $\Delta \nu_{\mathrm{fs,^ABe\rightarrow\,^9Be}}$ results in $^9$Be fine structure splitting with an accuracy of about 2.5\,ppm.  
	%Transfer of the available information from all isotopes to $^9$Be allows the determination of the fine structure spitting of this isotope with an accuracy of about 2.5\,ppm.
	}
	\label{fig:2}
\end{figure}

%\section {Theory}
The most convenient approach for the accurate description of light few-electron systems 
is based on NRQED. Relativistic, retardation, electron self-interaction, 
and vacuum polarization contributions can all be accounted for perturbatively 
by the expansion of the level energy in powers of the fine structure constant $\alpha$,
\begin{equation}
E(\alpha) = m\,\alpha^2\,{\cal E}^{(2)} + m\,\alpha^4\,{\cal E}^{(4)} + 
m\,\alpha^5\,{\cal E}^{(5)} + m\,\alpha^6\,{\cal E}^{(6)} + \ldots \label{01}
\end{equation} 
where the expansion coefficients ${\cal E}^{(i)}$ may include powers of $\ln\alpha$. 
The accuracy achieved for He-, Li-, and Be-like systems far exceeds
all other computational approaches, which rely on Dirac-like methods. 
It is particularly visible for the fine structure,
where relativistic methods such as RMBPT, RCC, MCDF, or RCI  
\cite{safronovaBeplus,dereviankoCC,pergerMCDF,yerokhinLilike}
have achieved only one or two significant digits, while the NRQED approach can provide 
about eight digits, \textit{e.g.}, in the helium fine structure of $P$ levels \cite{yerkrp}.

Here we extend results obtained for the fine structure in He and Li 
to Be$^+$. Since expansion coefficients ${\cal E}^{(i)}$ are expressed 
in terms of first- and second-order matrix elements of operators
with the nonrelativistic wave function, accuracy of the numerical calculation
strongly depends on the quality of this function. For example, Multiconfigurational Dirac-Fock calculations
\cite{blundellMCHF,fischerTBL,godefroidMCHF} are accurate only to three digits
because the wave function is a combination of Slater determinants and does not 
satisfy the cusp condition.
A much more accurate nonrelativistic wave function can be obtained using an explicitly
correlated basis such as Hylleraas functions \cite{lit_fine_yd,Yan2008,wangHeminus,Puc2009}.
Even though three-electron integrals with explicitly correlated functions 
are much more complicated than two-electron ones, the obtained numerical results for Be$^+$
can be almost as accurate as for He. We will report details in a separate paper \cite{Puc2015}. 
Numerical results are summarized in Table\,\ref{tab:Theo}. The fine structure arises at the order $m \alpha^4$, the nuclear recoil term at this order, denoted $E_{\rm fs}^{(4,1)}$, is comparable in size with $m\alpha^6$ contributions but of opposite sign. Finally, $E_{\rm fs log}^{(7,0)}$ are leading logarithmic contributions in $m\alpha^7$ and uncertainty due to uncalculated nonlogarithmic terms is estimated as 50\% of its size. The nuclear spin of odd isotopes leads to hyperfine-induced fine structure splitting, changing the splitting by $\delta E_{\rm fs}$. Accurate values reported for all isotopes in \cite{Puc2009} are used in the following analysis.   
In total, the fine structure splitting in $^9$Be amounts to 197\,068.0(25)\,MHz, which is about 4.5\,MHz larger than the experimental value in Table\,\ref{tab:expfrequency} corresponding to about $1.5\sigma$ of the combined uncertainties. Previous experimental results included in Table\,\ref{tab:Theo} were more than an order of magnitude less precise. Since we have measured the fine structure splitting of the other isotopes as well, we have used this information to significantly improve the accuracy of the $^9$Be splitting as described below. 

\begin{table}[thb]
\renewcommand{\arraystretch}{1.0}
\caption{Fine structure splitting of the $2p$ states in $^9$Be$^+$ in units of MHz. $\delta E_{\rm fs}$ is the hyperfine mixing correction, $E_{\rm fs}^{(4,1)}$ the nuclear recoil term at order $m\alpha^4$. The uncertainty due to neglected terms is estimated to be 50\% of $E_{\rm fs log}^{(7,0)}$. }
\label{tab:Theo}
\begin{tabular}{r r@{.}l c}
\hline \hline
         & \multicolumn{2}{c}{$^9$Be$^+$} & {Ref.}\\
\hline
$E_{\rm fs}^{(4,0)}$            & 197\,039&15 (8)    & \cite{Puc2009} \\
$E_{\rm fs}^{(4,1)}$            &      -21&27       & \cite{Puc2009} \\
$E_{\rm fs}^{(6,0)}$            &       45&4 (4)   & 
\\
$E_{\rm fs log}^{(7,0)}$        &        4&6 (23)   & \\
$\delta E_{\rm fs}$             &        0&03       & \cite{Puc2009}\\
$E_{\rm fs}$(theo)              &  197\,068&0 (24)  & this work   \\[1ex]
$E_{\rm fs}$(theo)              & \multicolumn{2}{l}{197\,024 (150)}    & Yan {\it et al.} \cite{Yan2008} \\
% $E_{\rm fs}$(exp)               & 197\,144.         & Ralchenko {\it et al.}\cite{ralchenko} \\ NIST value result from Bollinger result
$E_{\rm fs}$(exp)               & \multicolumn{2}{l}{197\,150 (64)}     & Bollinger {\it et al.} \cite{bollinger1985} \\
$E_{\rm fs}$(exp)               & 197\,063&48 (52)   & This work \\
\hline\hline
\end{tabular}
\end{table}
%
%\section{SIS and Projected Fine Structure} 
%Changes of the fine structure splitting between isotopes, the splitting isotope shift, 
The SIS can be traced back to two contributions: mass-dependent terms in the fine structure Hamiltonian and hyperfine-induced mixing. While in \cite{Yan2008} only the former have been calculated, the influence of the hfs was included in \cite{Puc2009}.  
 
The SIS relative to the $^9$Be fine structure splitting is plotted in the left part of Figure~\ref{fig:2}. The lowest point plotted in green is the SIS between $^{12}$Be and $^{10}$Be which can be determined to the highest precision since hfs is absent. Filled circles represent corresponding theoretical results. % for the SIS. 
It is striking that all experimental results practically coincide with theory, much better than expected from the size of the error bar. This reflects probably an overestimation of our systematic uncertainties or a much better cancellation of the contributing effects than expected in our conservative estimate. The only noticeable deviation from theory is observed for $^7$Be ($0.5\sigma$). In this case, no data has been taken in run II due to lack of beam time.

Since the calculated SIS is accurate at a level that by far exceeds that of the experiment (uncertainty on the kHz level) we can use it to compare the results for the different isotopes and to reduce the uncertainty of the $^9$Be measurement. The fine structure splitting in $^9$Be can be calculated from the measured splitting of other isotopes according to
\begin{equation}
\Delta \nu_{\mathrm{fs,^ABe\rightarrow\,^9Be}} = \Delta \nu_{\rm fs}(^A{\rm Be}) - \delta \nu^{A,9}_{\mathrm{sis, Theory}}.
\label{eq:fs-transf} 
\end{equation}
Results are included in Table\,\ref{tab:expfrequency} and plotted in Fig.\,\ref{fig:2}. 
%The splitting as obtained from the different isotopes scatter only slightly. 
The average splitting and standard deviation for $^9$Be 
%and the standard deviation of the values (including the directly measured one) 
is $197\,063.47(53)$\,MHz. 
%Using the information from all isotopes allowed a reduction of the uncertainty for the $^9$Be fine structure splitting by another factor of 4. In total our measurement has now an uncertainty improved by more than two orders of magnitude compared to the best previous one reported in \cite{bollinger1985}. 
Uncertainty is thus further reduced fourfold and more than two orders of magnitude compared to literature 
%the previously reported one 
\cite{bollinger1985}.
 
In summary we have measured transition frequencies in the D1 and D2 lines of $^{7,9-12}$Be and determined the fine structure splittings. 
%In view of the former problem in the lithium fine structure, 
The SIS was extracted and excellent agreement with \textit{ab initio} calculations was found. This verifies that the mass effect in such calculations is well under control. Using the calculated SIS values, we were able to transfer the accuracy of the measurements of the spin-zero isotopes $^{10,12}$Be to $^9$Be, resulting in a fourfold improvement of the measurement accuracy to 2.5\,ppm. Agreement between experiment and theory is reasonable and constitutes one of the most precise tests of QED in many-electron systems.
 
\begin{acknowledgments}
This work was supported by the Helmholtz Association (VH-NG148), the German Ministry for Science and Education  (BMBF Contract No.\ 05P12RDCIC), the Helmholtz International Center for FAIR (HIC for FAIR) within the LOEWE program by the State of Hesse, the Max-Planck Society, the European Union 7$^\mathrm{th}$ Framework through ENSAR, and the BriX IAP Research Program No. P7/12 (Belgium). A.~K. acknowledges support from the Carl-Zeiss-Stiftung (AZ:21-0563-2.8/197/1). M. P. and K. P. acknowledge the support of NCN grant 2012/04/A/ST2/00105 and by PL-Grid Infrastructure.
\end{acknowledgments}


\begin{thebibliography}{25}

%Introduction
%------------
\bibitem{Borbely2009}
J.S. Borbely, M.C. George, L.D. Lombardi, M.Weel, D.W. Fitzakerley, and E.A. Hessels, Phys.\ Rev.\ A \textbf{79}, 060503(R) (2009).
\bibitem{Din1991}
T.P. Dinneen, N. Berrah-Mansour, H.G. Berry, L. Young, and R.C.~Pardo, Phys.\ Rev.\ Lett.\ \textbf{66}, 2859 (1991).
\bibitem{Tho1998}
J.K.~Thompson, D.J.H.~Howie, and E.G.~Myers, Phys.\ Rev.\ A \textbf{57}, 180 (1998). 
\bibitem{Mye1999}
E.G.~Myers, H.S.~Margolis, J.K.~Thompson, M.A.~Farmer, J.D.~Silver, and M.R.~Tarbutt, Phys.\ Rev.\ Lett.\ \textbf{82}, 4200 (1999).
\bibitem{Pac2010}
K.~Pachucki and V.A.~Yerokhin, Phys.\ Rev.\ Lett.\ \textbf{104}, 070403 (2010).
\bibitem{Puc2014}
M.~Puchalski and K.~Pachucki, Phys.\ Rev.\ Lett.\ \textbf{113}, 073004 (2014).
\bibitem{Nob2006}
G.A.~Noble, B.E.~Schultz, H.~Ming, and W.A.~van Wijngaarden, Phys.\ Rev.\ A \textbf{74}, 012502 (2006).
\bibitem{Noe2011}
W.~N\"ortersh\"auser \textit{et al.}, Phys. Rev. A  \textbf{83}, 012516 (2011). 
\bibitem{San2011}
C.J.~Sansonetti, C.E.~Simien, J.D.~Gillaspy, J.N.~Tan, S.M.~Brewer, R.C.~Brown, S.~Wu,
and J.V.~Porto, Phys.\ Rev.\ Lett.\ \textbf{107}, 023001 (2011).
\bibitem{Bro2013}
R.C.~Brown, S.~Wu, J.V.~Porto, C.J.~Sansonetti, C.E.~Simien, S.M.~Brewer, J.N.~Tan, and J.D.~Gillaspy, Phys.\ Rev.\ A \textbf{87}, 032504 (2013).
\bibitem{Volotka2014}
A.V.~Volotka, D.A.~Glazov, V.M.~Shabaev, I.I.~Tupitsyn, and G.~Plunien, Phys.\ Rev.\ Lett. \textbf{112}, 253004 (2014).
%\bibitem{Volotka2009}
%A.V.~Volotka, D.A.~Glazov, V.M.~Shabaev, I.I.~Tupitsyn, and G.~Plunien,  Phys. Rev. Lett. \textbf{103}, 033005 (2009).
\bibitem{Beiersdorfer2005}
P.~Beiersdorfer, H.~Chen, D.B.~Thorn, and E.~Tr\"abert, Phys.\ Rev.\ Lett. \textbf{95}, 233003 (2005).
\bibitem{Wagner2013}
A.~Wagner, S.~Sturm, F.~K\"ohler, D.A.~Glazov, A.V.~Volotka, G.~Plunien, W.~Quint, G.~Werth, V.M.~Shabaev, and K.~Blaum, Phys.\ Rev.\ Lett.\ \textbf{110}, 033003 (2013).
%\bibitem{Lochmann14}
%M.~Lochmann {\it et al.}, Phys.\ Rev.\ A  \textbf{90}, 030501 (2014).
%\bibitem{Ullmann15}
%J.~Ullmann {\it et al.}, J.\ Phys.\ B  in print (2015).
\bibitem{bollinger1985} 
J.J.~Bollinger, J.S.~Wells, D.J.~Wineland, and W.M.~Itano, Phys.\ Rev.\ A {\bf 31}, 2711 (1985).
\bibitem{Noe2009}
W.~N\"ortersh\"auser {\it et al.}, Phys.\ Rev.\ Lett.\ {\bf 102}, 062503 (2009).
\bibitem{lit_fine_yd} Z.-C. Yan and G.W.F. Drake, Phys. Rev. A {\bf 66}, 042504 (2002).

%Experimental
%------------
\bibitem{Bot2014}
B.~Botermann \textit{et al.}, Phys.\ Rev.\ Lett.\ {\bf 113}, 120405 (2014).
\bibitem{Din1989}
U.~Dinger, J.~Eberz, G.~Huber, R.~Menges, R.~Kirchner, O.~Klepper, T.~K\"uhl, and D.~Marx, Nucl.\ Phys.\ A {\bf 503}, 331 (1989).
\bibitem{Rei1999}
J.~Reichert, R.~Holzwarth, Th.~Udem, and T.W.~H\"ansch, Opt.\ Comm.\ {\bf 172}, 59 (1999).
\bibitem{Kri2012}
A.~Krieger {\it et al.}, Phys.\ Rev.\ Lett.\ {\bf 108}, 142501 (2012).
% leading order + finite mass + mixing (Li, Be+)
\bibitem{Puc2009} 
M.~Puchalski and K.~Pachucki, Phys.\ Rev.\ A {\bf 79}, 032510 (2009).
\bibitem{Zak2010}
M.~Zakova {\it et al.}, J.\ Phys.\ G {\bf 37}, 055107 (2010).

%Theory 19-
%------------
\bibitem{safronovaBeplus} U. I. Safronova, and M. S. Safronova , Phys. Rev. A {\bf 87}, 032502 (2013).
\bibitem{dereviankoCC} A. Derevianko, S. G. Porsev, and K. Beloy, Phys. Rev. A {\bf 78} , 010503 (R) (2008).
\bibitem{pergerMCDF} W. F. Perger, and B. P. Das, J. Phys.B: At. Mol. Phys. {\bf 20}, 665 (1987).
\bibitem{yerokhinLilike} V. A. Yerokhin, and A. Surzhykov, Phys. Rev. A  {\bf 86}, 042507 (2012).
\bibitem{yerkrp} K. Pachucki and V.A. Yerokhin, Phys. Rev. Lett. {\bf 104}, 070403 (2010).
% theoretical results MCHF
\bibitem{blundellMCHF} 
S.A.~Blundell, W.R.~Johnson, Z.W.~Liu, and J.~Sapirstein, Phys.\ Rev.\ A {\bf 40}, 2233 (1989).
\bibitem{fischerTBL} 
C.~Froese Fischer, M.~Saparov, G.~Gaigalas, and M.~Godefroid, At.\ Data Nucl.\ Data Tables {\bf 70}, 119 (1998).
\bibitem{godefroidMCHF}  
M.~Godefroid, C.~Froese Fischer, and P.~J\"onsson, J.\ Phys.\ B {\bf 34}, 1079 (2001).
\bibitem{Yan2008}
Z.~C.~Yan, W~N\"ortersh\"auser and G.~W.~F.~Drake, Phys.\ Rev.\ Lett.\ {\bf 100}, 243002 (2008); ibid. {\bf 102}, 249903(E) (2009).
\bibitem{wangHeminus} L. Wang, C. Li, Z.-C. Yan, and G.W.F. Drake, Phys. Rev. Lett. {\bf 113}, 263007 (2014).
\bibitem{Puc2015}
M.~Puchalski and K.~Pachucki, arXiv:1506.02462 [physics.atom-ph]. 

\end{thebibliography}
\end{document}